\documentclass[12pt]{iopart}

\usepackage{graphicx}

\begin{document}

\title{Highly versatile atomic micro traps generated by multifrequency magnetic field modulation}

\author{Ph.W. Courteille, B. Deh, J. Fort\'agh, A. G\"unther, S. Kraft, C. Marzok, S. Slama, and C. Zimmermann}
\address{Physikalisches Institut, Eberhard-Karls-Universit\"at T\"ubingen,
\\Auf der Morgenstelle 14, D-72076 T\"ubingen, Germany}

\date{\today}

\begin{abstract}
We propose the realization of custom-designed adiabatic potentials for cold atoms based on multimode radio frequency radiation in 
combination with static inhomogeneous magnetic fields. For example, the use of radio frequency combs gives rise to periodic potentials 
acting as gratings for cold atoms. In strong magnetic field gradients the lattice constant can be well below $1~\mu$m. By changing the 
frequencies of the comb in time the gratings can easily be propagated in space, which may prove useful for Bragg scattering atomic matter 
waves. Furthermore, almost arbitrarily shaped potential are possible such as disordered potentials on a scale of several $100~$nm or 
lattices with a spatially varying lattice constant. The potentials can be made state selective and, in the case of atomic mixtures, also 
species selective. This opens new perspectives for generating tailored quantum systems based on ultra cold single atoms or degenerate 
atomic and molecular quantum gases. 
\end{abstract}


\maketitle

\section{Introduction}
\label{SecIntroduction}

The availability of matter in a coherent state has renewed the interest in Bragg scattering with massive particles. Various types of 
gratings have been developped based on arrangements of two nearly degenerate laser beams \cite{Kozuma99,Stenger99} or microfabricated 
structures \cite{GuntherA05}. These techniques have drawbacks however: Optical lattices are not flexible in the sense that it is 
difficult to implement local variations or disorder, their geometry is frozen by the incident radiation of a few laser beams, and 
their lattice constant is limited by the wavelength of the employed lasers. On the other hand, microstructured gratings are unalterable 
by their design, and they are generally two-dimensional. 

In magnetic traps incident microwave or radio frequency radiation can have a strong impact on the trapping potential. This is due to the 
fact that the potential depends on the magnetic substate of the trapped atoms. The energy of this substate can be manipulated by admixing 
other substates via resonant radio frequency radiation. In inhomogeneous magnetic fields this coupling is local and leads, within a 
dressed states picture, to avoided level crossings \cite{Rubbmark81}. Atoms moving across an area where the coupling is strong, follow 
adiabatic potentials by avoiding the crossings. 

This feature is e.g.~used for evaporative cooling, where trapped magnetic states are coupled to untrapped states at certain distances 
from the trap center \cite{Hess86,Ketterle96} and for certain schemes of output coupling atom laser beams from trapped Bose-Einstein 
condensates \cite{Mewes97}. Atomic traps based on avoided crossings have been proposed by Zobay \textit{et al.} \cite{Zobay01,Zobay04} 
and pioneered by Perrin \textit{et al.} \cite{Perrin04}. Meanwhile double-well potentials have been built using this approach 
\cite{Schumm05}, and resonators for atoms have been demonstrated \cite{Bloch01}. 

In this paper we propose to generate periodic potentials for atoms by combining appropriate magnetic field gradients with the application 
of multimode radio frequency radiation. As compared to optical or microfabricated gratings, gratings based on radio frequency combs 
possess a number of advantages. They work for atoms with low- \textit{and} high-field seeking magnetic moments with a phase of the 
grating, which depends on the atomic internal state. Furthermore, the technique allows for an almost complete control over the spatial 
shape and the temporal evolution of the potentials. In particular, irregular patterns may be formed, and very small structures, only 
limited by the size of the technically feasible magnetic field gradient, may give rise to gratings with very small lattice constants. 
With properly shaped magnetic fields almost arbitrary $3D$ potential geometries are imaginable. It is even possible to design 
time-dependent shapes, e.g.~to shift, merge or split individual potential sites. The range of possible manipulations and applications 
seems inexhaustible. 

\bigskip

We begin this proposal with a quantitative description of the basic mechanism in Sec.~\ref{SecSimple}, where we suggest a very simple 
approximation for the adiabatic potentials, which holds when the frequency components are not too closely spaced. To demonstrate the 
versatility of the radio frequency technique, we discuss two examples. In Sec.~\ref{SecLattices} we focus on the case of a periodic 
potential (or grating). We estimate the smallest periodicity, which can be realized using radio frequency combs, discuss how to generate 
moving gratings in order to perform Bragg scattering experiments and how to use the phenomenon of Bloch oscillations to probe such a 
grating. 

The second example, discussed in Sec.~\ref{SecSquare}, is a way of modifying the shape of a magnetic trap using microwave radiation. In 
ordinary magnetic traps, atomic spin states with identical Zeeman shifts experience the same confining force regardless of the mass of 
the trapped particles. The possibility to design magnetic potentials with arbitrary shapes provides a useful handle to selectively 
manipulate mixtures of different species.

\section{Simple theoretical description}
\label{SecSimple}

We consider an atomic ground-state with hyperfine structure having the total spin $F$. The magnetic sublevels $m_F$ are split in an 
external magnetic field $B$ by an amount $\mu_Bg_Fm_FB$, where $g_F$ is the atomic $g$-factor of the hyperfine level. An irradiated 
linearly polarized radio frequency $\mathbf{B}_{rf}\cos(\omega t)$ couples the sublevels $|F,m_F\rangle\leftrightarrow|F,m_F'\rangle$ 
with $m_F'=m_F\pm1$, wherever it is close to resonance, provided the orientations of the radio frequency and the magnetic fields are 
orthogonal. The coupling strength is given by the Rabi frequency \cite{Ketterle96} 
	\begin{equation}\label{Eq01}
	\Omega=\frac{\mu_B g_F}{4\hbar}\left|\mathbf{B}_{rf}\times\hat{e}_B\right|\sqrt{F(F+1)-m_Fm_F'}~,
	\end{equation}
where $\hat{e}_B$ is the orientation of the local static magnetic field. 

Alternatively, a microwave frequency can be used to couple \textit{different hyperfine levels}. However, the essence of the effect can be 
studied in the coupled system $|\frac{1}{2},\frac{1}{2}\rangle\leftrightarrow|\frac{1}{2},-\frac{1}{2}\rangle$, on which we will focus 
in most parts of this paper. A generalization to multilevel systems $F>\frac{1}{2}$ is straightforward. The dressed states Hamiltonian 
of our two-level system is a $2\times2$ matrix, 
	\begin{equation}\label{Eq02}
	H(z)=\left(\begin{array}[c]{cc}\frac{1}{2}\mu_Bg_FB(z)-\frac{1}{2}\hbar\omega & \frac{1}{2}\hbar\Omega\\
	\frac{1}{2}\hbar\Omega & -\frac{1}{2}\mu_Bg_FB(z)+\frac{1}{2}\hbar\omega\end{array}\right)~.
	\end{equation}
For simplicity we have also assumed a one-dimensional geometry, $B=B(z)$, but it can easily be generalized to three dimensions. The 
eigenvalues of $H$ are 
	\begin{equation}\label{Eq03}
	E_{\pm}(z)=\pm\frac{1}{2}\sqrt{\hbar^2\Omega^2+\left[\mu_Bg_FB(z)-\hbar\omega\right]^2}~.
	\end{equation}
Far enough from resonance, $\hbar\Omega\ll\left|\mu_Bg_FB(z)-\hbar\omega\right|$, we get 
	\begin{equation}\label{Eq04}
	E_{\pm}(z)\approx\pm\frac{1}{2}\left[\mu_Bg_FB(z)-\hbar\omega\right]\pm\frac{\hbar^2\Omega^2}{4\left[\mu_Bg_FB(z)-\hbar\omega\right]}~,
	\end{equation}
where the second term can be interpreted as dynamic Stark shift of the energy levels. 
	
\begin{figure}[ht]
		\centerline{\scalebox{0.5}{\includegraphics{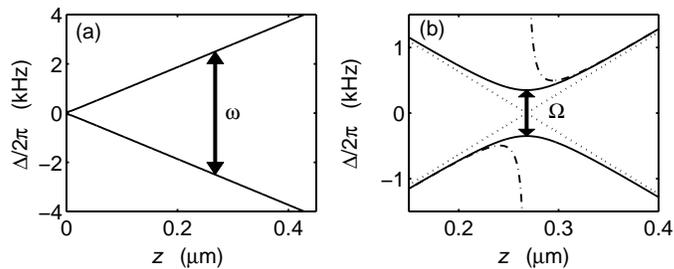}}}\caption{
			\textbf{(a)} Potential energies for a $F=\frac{1}{2}$ level scheme with the $g$-factor of $g_F=-\frac{2}{3}$ (as is the case 
				for the $^6$Li ground state $^2S_{1/2}$). A single radio frequency component (arrow) couples the substates 
				$m_F=\pm\frac{1}{2}$. Here $b=200~$G/cm and $\omega=2\pi\times 5~$kHz. 
			\textbf{(b)} Uncoupled dressed states (dotted line), coupled dressed states (solid), and dynamic Stark shifts (dash-dotted) 
			    calculated in the off-resonant field approximation. The Rabi frequency is $\Omega=2\pi\times 700~$Hz.}
		\label{Figure1}
		\end{figure}

To illustrate the action of the radio frequency, we calculate the potential energy and the dressed states for $^6$Li atoms. For 
simplicity, we assume a $1D$ linear magnetic field gradient $B(z)\equiv zb$. Fig.~\ref{Figure1}(a) shows the 
radio frequency coupling and Fig.~\ref{Figure1}(b) the dressed states for two magnetic substates coupled by a single 
radio frequency component. 

\begin{figure}[ht]
		\centerline{\scalebox{0.5}{\includegraphics{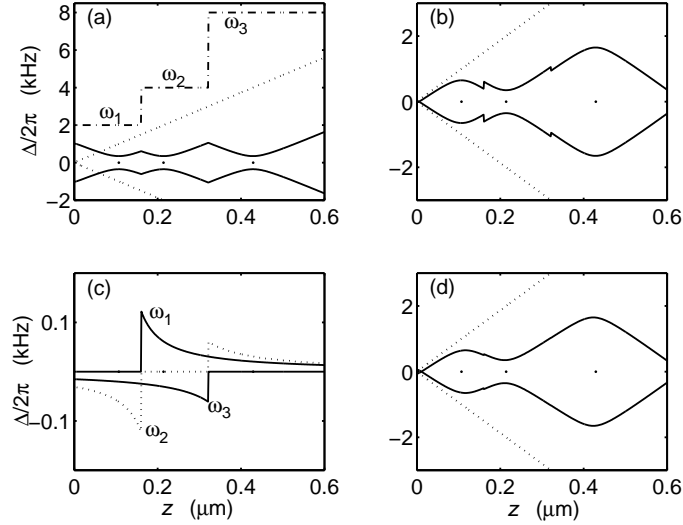}}}\caption{
			\textbf{(a)} Local dressed states (solid line). The parameters are the same as in Fig.~\ref{Figure1}, but with 
			    three frequencies irradiated at $\omega_n=2\pi\times 2,4,8~$kHz. The black dots show the locations where the frequencies 
			    are resonant. The dash-dotted line denotes the frequency component $\omega_n$, which is closest to resonance. 
			\textbf{(b)} Corresponding adiabatic potentials calculated with nearly resonant frequencies, only. 
			\textbf{(c)} Dynamic Stark shift corrections from those radio frequency fields, which are \textit{not} closest to resonance. 
			\textbf{(d)} Adiabatic potentials with the Stark shift corrections of Fig.~(c).}
		\label{Figure2}
		\end{figure}

If we irradiate several frequency components $\omega_n$, labelled in ascending order by $n=1,2,..$, the problem gets more complicated, 
because every component has its own dressed states basis. However for sufficiently low Rabi frequencies and large frequency separation 
of the components the dynamics is essentially governed by the component which is closest to resonance. A good approximation holding for 
homogeneous magnetic fields is therefore to consider \textit{only one} nearly resonant frequency and disregard the others. To handle 
the case of magnetic field gradients, we introduce a \textit{local} frequency, which at a given location $z$ is the frequency component 
closest to resonance, $\omega(z)=\omega_{n(z)}$, where the component $n=n(z)$ is chosen such that $|\mu_Bg_FB(z)-\hbar\omega_{n(z)}|$ is 
smallest. In other words, along the field gradient we switch between different dressed state representations; and in order to make this 
switching as smooth as possible, it is done at those points where a blue-detuned frequency $\omega_{n+1}$ gets closer to resonance than 
a red-detuned frequency $\omega_n$. Fig.~\ref{Figure2}(a) reproduces the energies locally dressed with the radio frequency field 
$\omega_n$. 

The adiabatic potentials $V_{ad}$ are now obtained by considering that the couplings are strong enough to yield Landau-Zener transition 
probabilities close to unity (see below). Formally, this is done using the following procedure: 
	\begin{equation}\label{Eq05}
    V_{ad,\pm}(z)=(-1)^{n(z)}\left[E_{\pm}(z) \mp \frac{\hbar\omega_{n(z)}}{2}\right] \mp \sum_{k=1}^{n(z)-1}~(-1)^k\hbar\omega_k~.
	\end{equation}
This is shown in Fig.~\ref{Figure2}(b). 

The adiabatic potentials in Fig.~\ref{Figure2}(b) show small discontinuities at those places, where the frequency component 
which are nearest to resonance changes. This is an artefact due to our neglecting of all non-resonant components. The error made can be 
reduced by considering their combined Stark shifts 
	\begin{equation}\label{Eq06}
	L_n(z)\equiv\sum\limits_{j\ne n}\frac{\hbar^2\Omega^2}{4\left[\mu_Bg_FB(z)-\hbar\omega_j(z)\right]}
	\end{equation}
in the \textit{diagonal} components of the Hamiltonian 
	\begin{equation}\label{Eq07}
	H_{\omega}(z)=H_{\omega_n}(z)+L_n(z)\left(\begin{array}[c]{cc} -1 & 0 \\ 0 & 1 \end{array}\right)~.
	\end{equation}
The corrected eigenfrequencies are thus 
	\begin{equation}\label{Eq08}
	E_{\pm}(z)=\pm\frac{1}{2}\sqrt{\hbar^2\Omega^2+\left[\mu_Bg_FB(z)-\hbar\omega+2L_n(z)\right]^2}~. 
	\end{equation}
Fig.~\ref{Figure2}(c) visualizes the Stark shifts produced by those frequency components, which are not the closest to resonance. 
Recalculating the adiabatic potentials~(\ref{Eq05}) with the corrected eigenfrequencies we obtain potentials without discontinuities as 
shown in Fig.~\ref{Figure2}(d). 

\bigskip

A major concern is the dependence of the coupling strength on the relative orientation of the radio frequency polarization and the 
magnetic field. Any variation of the field vector orientation over the region, where the atoms are trapped, introduces a 
position-dependence of the Rabi frequency. This problem can be avoided by chosing a sufficiently large magnetic offset field, which  
defines a spatially invariant quantization axis, and by arranging for an orthogonal polarization vector. On the other hand, this feature 
may be technically exploited as an additional handle for shaping the adiabatic potentials \cite{Schumm05}. 

\bigskip

\begin{figure}[ht]
		\centerline{\scalebox{0.48}{\includegraphics{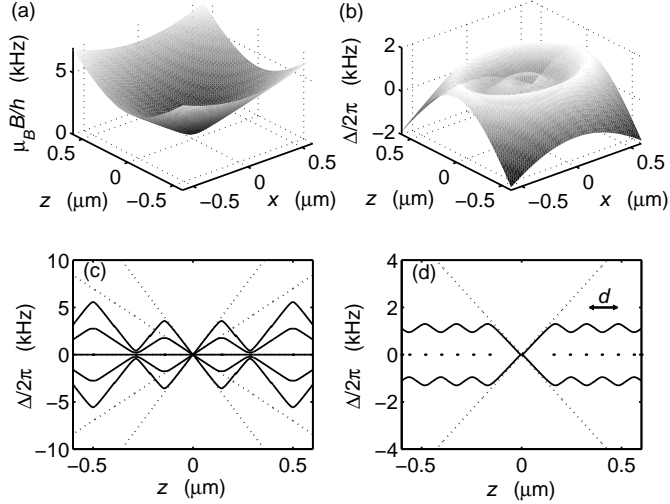}}}\caption{
			\textbf{(a)} $2D$ quadrupolar magnetic field with gradients $b_z=200~$G/cm and $b_x=140~$G/cm. 
			\textbf{(b)} Corresponding adiabatic potentials for the $^6$Li two-level system $F=\frac{1}{2}$. The Rabi- and radio 
				frequencies are $\Omega=2\pi\times 500~$Hz and $\omega_n=2\pi\times 2,4,7~$kHz, respectively. 
			\textbf{(c)} Adiabatic potentials for a system with five Zeeman substates, $F=2$. The $g$-factor is set to $g_F=\frac{1}{2}$, 
				as is the case for the $^7$Li ground state $^2S_{1/2}$. The Rabi- and radio frequencies are as in Fig.~(b).
			\textbf{(d)} Adiabatic potentials for a frequency comb and the $^6$Li two-level system. In this case the Rabi frequency is 
				chosen to be $\Omega=2\pi\times 400~$Hz.}
		\label{Figure3}
		\end{figure}

The above procedure can easily be extended to an arbitrary number of frequency components, to higher-dimensional magnetic field shapes 
$\mathbf{B(r)}$, or to multilevel systems $F>\frac{1}{2}$. Fig.~\ref{Figure3}(a,b) visualizes the case of a two-dimensional 
quadrupolar magnetic field $|\mathbf{B}(x,z)|=\sqrt{(z\nabla_zB)^2+(x\nabla_x B)^2}$ with three frequencies irradiated. 
Fig.~\ref{Figure3}(c) shows the adiabatic 
potentials for five Zeeman substates, as in the case of the $^7$Li ground state hyperfine level $F=2$. Finally, Fig.~\ref{Figure3}(d) 
represents the adiabatic potentials for the case of a frequency comb $\omega_n=2\pi\times(3+1.5n)~$kHz. 
Note that gravity is easily taken into account by calculating the resonant frequencies as explained above and supplementing the 
Hamiltonian (\ref{Eq07}) with the potential energy $E_{pot}=mgz$.

\section{Radio frequency lattices}
\label{SecLattices}

We have seen above that by irradiating a frequency comb we obtain a periodic potential [cf. Fig.~\ref{Figure3}(d)]. The 
question arises how small the lattice constant can be made. The periodicity depends on the spacing of the frequency components 
and on the magnetic field gradient. The grating is characterized by the lattice constant  
	\begin{equation}\label{Eq09}
    d=\frac{2\hbar(\omega_{n+1}-\omega_n)}{\mu_Bg_Fb}~,
    \end{equation}
and the potential modulation depth 
	\begin{equation}\label{Eq10}
    V_{ad}=\hbar\left(\frac{\omega_{n+1}-\omega_n}{2}-\Omega\right)~.
    \end{equation}
To obtain small lattice constant, we chose large gradients. For the following, we set the premise that the maximum feasible gradient is 
$b=200~$G/cm, which can be achieved with current technologies over reasonably large spatial ranges. 

\begin{figure}[ht]
		\centerline{\scalebox{0.5}{\includegraphics{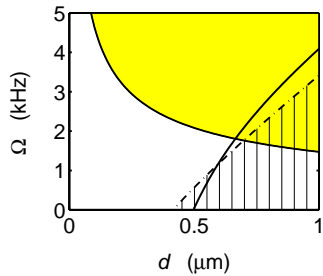}}}\caption{
			The shaded area shows the adiabatic regime for $\Omega$ and $d$ with a field gradient of $b=200~$G/cm. The zone at the 
				left of the solid line corresponds to gratings shallower than one recoil energy. The dash-dotted line delimits the 
				interband tunneling regime from the adiabatic regime (hatched area) as explained in the text.}
		\label{Figure4}
		\end{figure}

A typical experiment with gratings consists in Bragg-scattering moving atoms. Atoms with a de Broglie wavelength which matches the 
lattice periodicity are Bragg-reflected. The corresponding atomic velocity is 
	\begin{equation}\label{Eq11}
	v=\frac{h}{md}~.
	\end{equation}
The adiabaticity criterion sets now an upper limit for the velocity quantified by the probability for Landau-Zener transitions, 
$P_{LZ}=1-e^{-h\Omega^2/\partial_t\left[\mu_Bg_FB(z)\right]}$ with $z=vt$. The atoms follow the adiabatic potentials if 
$P_{LZ}\rightarrow1$. This implies $h\Omega^2\gg\mu_Bg_Fvb$. Consequently, in order to ensure Landau-Zener transitions for velocities as 
high as the Bragg velocity, the frequency comb spacing and the Rabi frequency should be chosen such that 
	\begin{equation}\label{Eq12}
	m\Omega^2d\gg\mu_Bg_Fb~.
	\end{equation}
The shaded zone in Fig.~\ref{Figure4} depicts the adiabatic regime. 

The interplay of the parameters $\Omega$ and $d$ also governs the potential modulation depth $V_{ad}$ through the Eqs.~(\ref{Eq09}) and 
(\ref{Eq10}). This relationship may be illustrated by estimating the smallest lattice constant, for which the lattice potential is still 
deeper than one recoil energy, $E_r=h^2/8md^2$, as a function of $\Omega$. The solid line in Fig.~\ref{Figure4} reproduces the 
curve $d=d(\Omega)$ for which $V_{ad}=E_r$. We find that for a Rabi frequency of $3~$kHz the lattice constant must exceed 
$d\approx0.7~\mu$m. 

The lattice constant can be reduced using larger gradients. Lattice structures on a length scale of a few $100~$nm, which are certainly 
feasible, could be of interest for the generation of disordered potentials with the aim of studying atomic Bose and Anderson glasses 
\cite{Damski03}. The most interesting regime, where the length scale of the disorder is smaller than the healing length of a 
Bose-Einstein condensate, is difficult to reach with optical speckle patterns \cite{Lye05,Clement05}. In contrast the regime is quite 
accessible to radio frequency lattices.

\subsection{Moving gratings} 

\begin{figure}[ht]
		\centerline{\scalebox{0.48}{\includegraphics{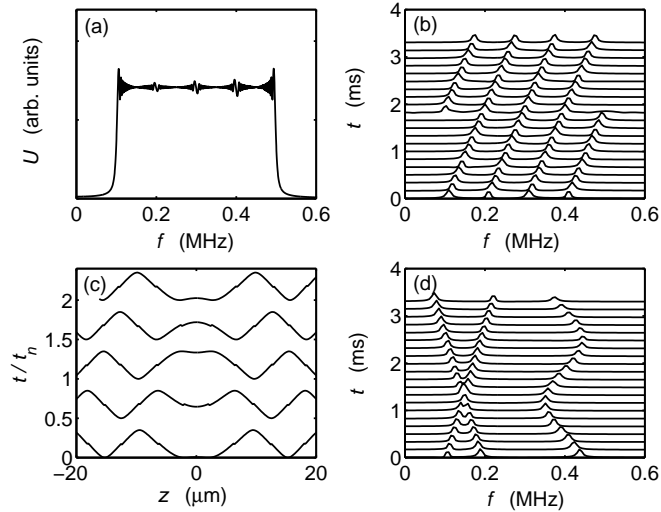}}}\caption{
			\textbf{(a)} Overall Fourier transform of the signal of 4 radio frequency components linearly scanned across a range 
				corresponding to their frequency splitting. 
			\textbf{(b)} Stepwise Fourier transform of the same signal. 
			\textbf{(c)} Time-evolution of corresponding adiabatic potentials. The gradient is again $b=200~$G/cm. Additionally, a 
				magnetic offset field of $B_0=0.1~$G has been applied to avoid Majorana spin flip transition at the center of the quadrupole 
				trap. The Rabi frequency is $\Omega=2\pi\times 15~$kHz. 
			\textbf{(d)} Stepwise Fourier transform of a signal consisting of three frequency components, the amplitudes and phases of 
				which vary differently in time.}
		\label{Figure5}
		\end{figure}

We now want to produce an adiabatic grating which propagates in a constant magnetic field gradient. This necessitates a time-dependent 
frequency comb, $\omega_n(t)$. After a time $t_n$ the frequency drift has covered the separation of the frequency components 
$\omega_{n+1}(t_n)=\omega_n(0)$, and we recover the same spectrum. The frequency separation determines the lattice constant through 
Eq.~(\ref{Eq09}). The time $t_n$ rules the propagation velocity of the lattice via 
	\begin{equation}\label{Eq13}
	v=d/t_n~.
	\end{equation}
Therefore we may repeat the ramp 
	\begin{equation}\label{Eq14}
	\omega_n(t)=\omega_n(0)+(\omega_{n+1}-\omega_n)~t_n^{-1}~(t~\mathrm{mod}~t_n)~.
	\end{equation}
The Fourier transform of the \textit{instantaneous} power spectrum of the radio frequency, 
$P(\omega,t)=\sum_n\delta[\omega-\omega_n(t)]$, is 
	\begin{equation}\label{Eq15}
	U(\tau,t)=U_0\sum_n{\sin[\omega_n(t)\tau]}~.
	\end{equation}
The inverse Fourier transform of the function $U(t,t)$, shown in Fig.~\ref{Figure5}(a), is the envelope of the frequency comb 
time evolution. To retrieve the instantaneous spectrum, we build the invers Fourier transforms over consecutive time intervals as shown 
in Fig.~\ref{Figure5}(b). 

In the example of Fig.~\ref{Figure5}, the chosen time base is $t_n=2~$ms. In a magnetic field gradient $b=200~$G/cm, 
with a frequency comb chosen as $\omega_n(0)=2\pi n\times100~$kHz, the lattice constant is $d=1.7~\mu$m. Thus the propagation velocity is 
$v=1.8~$mm/s. 

\bigskip

Multifrequency signals can be generated by a variety of techniques. The most convenient is probably to use arbitrary waveform generators 
programmed with the Fourier transform of the desired frequency spectrum. In order to generate moving radio frequency lattices, a whole 
period $t_n$ has to be programmed. In our example, for radio frequencies up to $\omega_n\approx2\pi\times 500~$kHz the number of 
oscillations within $t_n$ is $1000$. If we count at least $8$ points to resolve an oscillation, we need to program waveforms with $8000$ 
points, which is feasible with state-of-the-art arbitrary waveform generators.

\subsection{Bragg reflection and Bloch oscillations}

Atoms are Bragg-reflected at a lattice if their de Broglie wavelength is commensurable with the lattice constant, i.e.~their velocity 
is given by Eq.~(\ref{Eq11}). Alternatively, for resting atoms to be accelerated by a moving grating, the propagation velocity of the 
grating Eq.~(\ref{Eq13}) must coincide with the Bragg velocity. For the above parameters, we expect $v_{brg}=1.2~$mm/s. For our example, 
in order to obtain an interaction between the atoms and the lattice, we may tune the propagation velocity by varying the time period 
$t_n$ or modifying the lattice constant $d$. In contrast to the propagation velocity, which increases with the lattice constant, the 
resonant Bragg velocity is inversely proportional to it. Hence, there can only be one field gradient satisfying the Bragg condition. 
Equating the two velocities, $v=v_{brg}$, we derive 
	\begin{equation}\label{Eq16}
	b_{brg}=\sqrt{\frac{2\hbar m}{\pi t_n}}\frac{\omega_{n+1}-\omega_n}{\mu_Bg_F}~,
	\end{equation} 
which gives $b_{brg}\approx210~$G/cm for the present conditions. 

Thus for anisotropic traps, for which the secular frequencies differ for different axis, a pronounced directional selectivity of the 
atomic interaction with the moving lattice can be expected. The directions in which atoms are accelerated are only those, where the 
Bragg condition is satisfied. 

Fig.~\ref{Figure5}(c) shows a time series of the adiabatic potentials produced by the frequency comb specified above. When the frequency 
components are ramped, the potential wells at both sides of the trap origin migrate in opposite directions. This feature can be useful 
for realizing atom interferometers. 

The technique is obviously not limited to uniform gratings propagating with constant speed. As an example that arbitrary time signals 
may be programmed Fig.~\ref{Figure5}(d) shows the sequential Fourier transform of a waveform consistent of three frequency 
components, which are differently modulated in amplitude and phase. 

\bigskip

A powerful tool to probe and characterize atom optical gratings are \textit{Bloch oscillations}. This phenomenon occurs when a constant 
force, $F=ma$, is applied to cold atoms interacting with the grating. The force can be gravity, or it can be generated by an accelerated 
motion of the grating \cite{BenDahan96,Peik97}. For Bloch oscillations to occur in lowest band, an adiabaticity criterion similar to the 
Landau-Zener criterion (\ref{Eq12}) follows from the Bloch model: Interband tunneling at the edge of a Brillouin zone is prevented if 
the rate of change of potential energy due to acceleration is smaller than the size $V_{ad}$ of the band gap, 
i.e.~$V_{ad}^2\gg\frac{d}{dt}\left.(\hbar maz)\right|_{z=d/2}$. This leads immediately to 
	\begin{equation}\label{Eq17}
	\frac{\pi}{4}\frac{V_{ad}^2}{E_r}\gg mad~.
	\end{equation}
Inserting the recoil energy and the expressions~(\ref{Eq09}) and (\ref{Eq10}), we obtain an inequality for $\Omega$ and $d$. This 
inequality is depicted in Fig.~\ref{Figure4} as the hatched zone below the dash-dotted curve for the case that the acceleration 
is due to gravity $a=9.81~$m/s.

\section{Potential shaping}
\label{SecSquare}

Another application of adiabatic potentials is the realization of non-harmonic potentials via the irradiation of appropriate radio 
frequencies and microwaves. Two-dimensional trapping potential based on this principle have recently been proposed 
\cite{Zobay01,Zobay04}. The possibility to selectively influence the confinement strength of atoms trapped in specific Zeeman substates 
or being of different species is particularly useful for studies of their interaction. 

An example where this plays a role is a mixture of the high-field seeking states $^6$Li $|\frac{3}{2},\frac{3}{2}\rangle$ and $^{87}$Rb 
$|2,2\rangle$ \cite{Silber05}. For this case, sympathetic cooling of Li via Rb to very low temperatures (such as those needed for 
reaching the critical temperature for BCS-pairing) requires a sensitive matching of the respective gas densities: This guarantees that 
the Li Fermi temperature is lower than the critical temperature for Bose-Einstein condensation of Rb \cite{Note01} and reduces 
fermion-hole heating as far as possible \cite{Cote05}. This respective density-matching is now possible by controlling the potentials' 
shape via microwave or radio frequencies. 

\begin{figure}[ht]
		\centerline{\scalebox{0.49}{\includegraphics{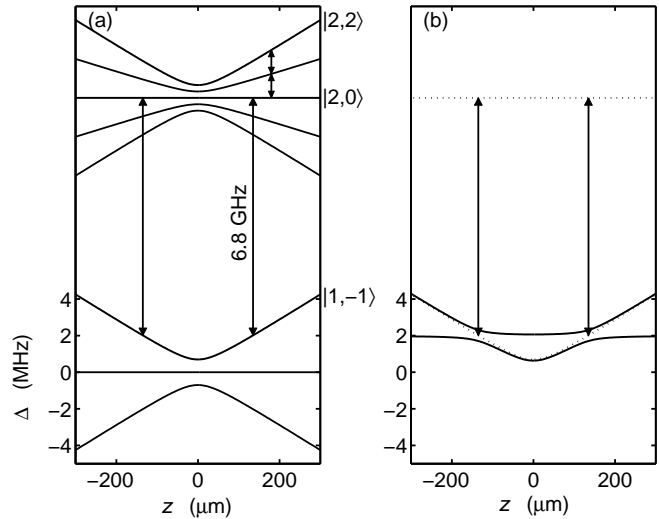}}}\caption{
			\textbf{(a)} Hyperfine and Zeeman structure of the $^{87}$Rb ground state. States $|2,0\rangle$ and $|1,-1\rangle$ are 
				coupled by a microwave at $6.8~$GHz. The Rabi frequency is $\Omega=2\pi\times 600~$kHz, the gradient is $b=200~$G/cm, 
				and we assume a magnetic offset field of $B_0=1~$G. The microwave is tuned $2~$MHz to the red of the field-free 
				$|1,-1\rangle\leftrightarrow|2,0\rangle$ resonance. 
			\textbf{(b)} Dressed states give rise to two adiabatic potentials. The upper potential is flat around the center, the lower 
				one has a depth of about $50~\mu$K for the chosen set of parameters.}
		\label{Figure6}
		\end{figure}

For this example, the following scenario could be used to transfer atoms into a potential with a flat bottom in which an atomic cloud 
adopts a nearly homogeneous density distribution. One starts producing an atomic cloud, e.g.~of $^{87}$Rb atoms in the $|2,2\rangle$ 
state. With a properly designed adiabatic sweep the majority of the atoms can be transferred to $|2,0\rangle$, and a microwave frequency 
resonant to the $|1,-1\rangle$ state at some distance from the trap center is irradiated [cf. Fig.~\ref{Figure6}(a)] \cite{Note02}. 
The atoms are then trapped in an adiabatic potential with a flat bottom [cf. Fig.~\ref{Figure6}(b)]. Note that there is a second 
trapping potential for those $|2,2\rangle$ atoms that were initially in the center. 

\section{Conclusion}
\label{SecConclusion}

To conclude we believe that radio frequency combs in conjunction with magnetic field are a versatile tool for manipulating the atomic 
motion on small length scales. The gratings are very stable and flexible. In fact the lattice constant is only limited by the magnetic 
field gradient, which is technically feasible. In our examples gradients of $200~$G/cm have been used, but gradients of several 
$1000~$G/cm can certainly be realized in environments, which are compatible with cold atom optics. Potential applications of time- or 
position-dependent radio frequency combs may include Bragg velocity filters, Bragg interferometers, quasi-random potentials with 
disorder occurring on a very small length scale, and other atom optical elements. 

In a variety of cases, such as in two-species experiments, it is important to control the atomic densities without modifying the 
magnetic field gradients. As an example, we have discussed the $^6$Li - $^{87}$Rb mixture, where the feature of deforming magnetic 
trapping potentials by radio frequency radiation in a controlled manner is crucial in order for sympathtic cooling to be efficient down 
to very low temperatuires. This approach may thus constitute an alternative to using bichromatic optical dipole traps suggested in 
Ref.~\cite{Presilla03}. 

\bigskip

We acknowledge financial support from the Deutsche Forschungsgemeinschaft (Zi419/4) and the European Union (MRTN-CT-2003-505032).

\end{document}